# Tamil Open-Source Landscape – Opportunities and Challenges

*Muthiah Annamalai[*], T. Shrinivasan[+]*

* - ezhillang@gmail.com, + tshrinivasan@gmail.com

## Introduction

General tenet of Open-Source and FreeSoftware originally founded by pioneers like Richard Stallman and Eric. S. Raymond [1a,b] continues to rely on collective good of developing software in open and creatively monetizing it outside of closed-source traditional models. Tamil open-source software has its roots from various GNU/Linux user-groups across Tamilnadu [2a], and individuals motivated by Tamil support on Internet [2b,c]. More recently a rise in awareness created by Tamilnadu branch of FSF [2d].

Tamil open-source software (TOSS) continues to grow with over a hundred repositories in github that contribute code for libraries, applications, speech synthesis tools, Android/iOS apps, OCR tools, web-utilities, translations and font faces. While Tamil open-source work truly may have started with translation efforts and localization of KDE (tamillinux in early 2000s) and GNOME (especially GTK) in same period of late 90s and early 2000s, today it continues to grow in an organic, if unorganized way. In the meanwhile many projects, have come and gone and morphed in their existence. (Please note this is not, by any means complete, representative, inclusive history of Tamil open-source development or Tamil computing – just a partial glimpse).

The various challenges are identified below, will be elaborated with appropriate examples,
1. Limited market for Tamil software
2. Marketing efforts for digital Tamil products
3. Lack of reusable, ready s/w components for Tamil software delays development and increases costs of development, production and post-production are limiting future projects
4. Tamil origin Tech workers are indifferent to TOSS causes

Barriers to entry into the Tamil open-source software (TOSS) space are identified and removal of these issues could spur a growth phase in TOSS
1. Software not addressing the market – teaching developers to address real market needs
2. Address demographic needs:
    What about other adults, young-adults, teenagers, boys and girls usage of software?
3. Sometimes CS challenges are hard – CS education and continuous growth are recommended for developers
4. Multiple roles required for software development, graphic art, testing, documentation, packaging and release, which are lesser known to potential Tamil contributors

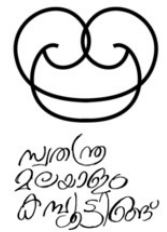

*Image 1: Swatantra Malayalam Computing*

We also like to highlight several open-source Tamil projects which helped create newer projects and grow the ecosystem. We also found Tamil project "OCRWikisource" by one of authors inspired an Oriya language application. So there is tremendous potential for growth in Indian-language computing by learning from one-another.

Further issues in TOSS include leadership and fund-raising abilities, continuity of the developer community sustenance and growth, promotion of open-source way of contributing to digital Tamil. For security of the Tamil open-source community we realize that a non-profit model with Wikipedia style rapid-grants for 6month periods to grow the community, certify young engineers contributions, and provide leadership will be ideal. Currently Thamizha and Ezhil Language Foundation have provided some aspect discussed here. Malayalam engineers have provided leadership for their efforts through Swatantra Computing [3a] (SMC) and maintaining frameworks like SILPA and 11-Indian language TTS system called Dhvani [3b].

We provide a list of representative Tamil open-source projects and number of contributors, and bugs reported/fixed and the pull requests from GitHub data in Table. 1. Our full review of TOSS landscape is expected to serve as a milestone of present day Tamil adoption and growth aspects.

| Project | Stars | Language | closed issues | open issues | pull requests | Fork | Contributors |
|---|---|---|---|---|---|---|---|
| **Ezhil-Lang** | 78 | Python | 100 | 93 | 59 | 35 | 11 |
| **open-tamil** | 37 | Python, Java, | 61 | 54 | 8 | 26 | 6 |
| OCR4wikisource | 26 | Python | 54 | 32 | 9 | 17 | 5 |
| **EkType/Mukta** | 46 | Font | 24 | 2 | 5 | 16 | 3 |
| **ratreya/lipika-ime** | 21 | Objective-C | 17 | 3 | 9 | 13 | 2 |
| **VanillaandCream/Catamaran-Tamil** | 26 | Font | 7 | 10 | 4 | 7 | 3 |
| FreeTamilEbooks (Android client) | 6 | Android/Java | 3 | 10 | 0 | 11 | 1 |
| thamizha/eKalappai | 11 | C++ | 3 | 13 | 4 | 8 | 3 |
| **velsubra/Tamil** | 6 | Java | 2 | 0 | 5 | 3 | 2 |
| ashokr/TamilNLP | 9 | Python | 1 | 0 | 1 | 4 | 1 |
| **mayooresan/Android-TamilUtil** | 15 | Java | 0 | 0 | 0 | 15 | 1 |
| **thamizha/android-tamilvisai** | 14 | Java | 0 | 3 | 3 | 14 | 6 |
| **godlytalias/Bible-Database** | 18 | PL/pgSQL | 0 | 0 | 1 | 9 | 2 |
| **vasurenganathan/tamil-tts** | 24 | php | 0 | 0 | 0 | 6 | 1 |
| **rdamodharan/tamil-stemmer** | 22 | C | 0 | 0 | 0 | 4 | 1 |
| echeran/CLJ-Thamil | 36 | Clojure | 0 | 0 | 0 | 3 | 1 |
| psankar/Korkai | 10 | Go | 0 | 0 | 0 | 3 | 1 |
| rprabhu/Tamil Dictionary | 8 | Javascript | 0 | 0 | 0 | 2 | 1 |

**Table 1:** Representative List of Tamil Open-Source projects by Git-Hub community

### GitHub collaboration space

GitHub is an open-source project development space that is popular and many Tamil origin developers are present and contributing to various open-source efforts – within and outside of Tamil computing.

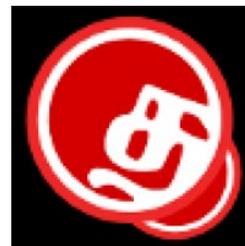

*Image 2: Thamizha group at Git Hub and http://thamizha.org*

Typically the project founders create a GitHub repository and choose one of the open-source licenses like MIT, Apache, GPL etc. and start uploading their code and setting up unit-tests and continuous-integration tests via Travis-CI. Other developers may join in by forking the git repository and working on one or more issues (bugs) or features and sending a pull-request to the original (founder's) repository.

The founder can have one or more comments and after resolving any outstanding disagreements the software pull-request is committed to the source repository and merged into one. Rinse-and-repeat of this flow is usual Git-Hub development

### Collectives in GitHub

Thamziha group [4] in GitHub is the single largest pool of volunteer developers followed behind by Ezhil-Language-Foundation. Thamizha group has 42 volunteer developers, 24 source projects, 2 forks, and contains sources for important projects like eKalappai, tamil-fonts, visaineri, Peyar, and Thamizha.org website. Major coding languages used in 26 repositories of Thamizha are JavaScript, Python, HTM, PHP. and Java. Thamizha is a open-source volunteer group with a supporting mailing list at freetamilcomputing [5], with -the logo in Image. 1.

## Repositories by Language

To gain a measure of interest and popularity of language based projects by developers in Github, we collected data by running searches on Github [7] and found the surprising result in Image. 3. Tamil language repositories outnumber Hindi or Malayalam, Kannada and Telugu efforts together.

While our data collection methods are by Github search and admittedly coarse grained, we still see on the 760 Tamil projects hosted on Github, even if only 20% of them are active, a resounding interest in Tamil computing. We need all of these 760 projects and developers to come on to mainstream and become active, thriving contributors. This data is a call for general Tamil community to support these budding open-source developers and channelize their energies to create useful, inventive software to further improve Tamil computing landscape. The more active among these 760 projects are dictionaries, Android and iOS applications, keyboards, transliteration software, programming languages in Tamil, encoders/converters for Tamil, Tamil font collections, book translations etc.

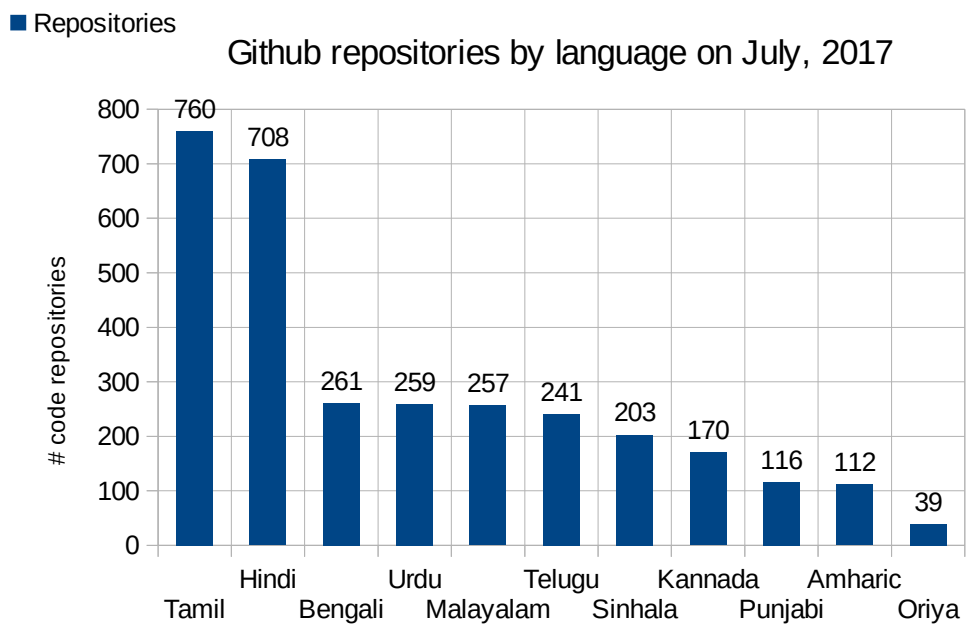

*Image 3: Chart of repositories in Github by language in July, 2017.*

# Why TOSS does not reach end-users?

With such a large body of software available within reach of developers it is natural to wonder why Tamil software usage has not "taken-off" in a big way. In this section we present some detailed analysis of why Tamil open-source software has missed the big-impact on general Tamil users and continues to under-perform its reach.

We present this analysis naming several valuable, popular projects with intention of highlighting their missing pieces and improving their impact – with complete understanding of how strapped for support, funding and manpower many of these projects can be. Hence authors request creative license to critique:

1. Software is not packaged well
   Most of the software are in development stage always; except a handful of TOSS projects rest of them binary executable packages for Linux/Windows is available. Not everyone has access to a development environment. This batteries excluded approach fails us.

   Examples:
   (a) Until recently the Ezhil-Language software was not available for downloads to Windows and Linux platform for general users to try out.
   (b) A popular software component often requested by developers but not available as a component is the tamil-stemmer by R. Damodharan [8]

2. Lack of online demos
   Even for software libraries, there no online demonstrations to have a taste of them, before installing. Example - open-tamil, has a non-unicode to unicode convertor, with many font combinations then any other available tool. since, it has no web interface or site, nobody knows and uses it.

3. No windows version

   As windows is still used by regular users, we need to ship packagesfor windows too. Example : OCR4WikiSource is a connector for Google OCR and Wikisource, but works only for Linux. New users found it difficult use with Linux. So, they not using it.

4. No showcase site for all the web applications or fonts

   There are many web applications for Tamil. But there is no showcase site to list them all, install and let the users to play around it. Few apps have their own sites. But they work sometime and die without maintenance.

   We have TamilTTS, avalokitam, Velmurugan's Tamil tools [9], anunaadam, Pallanguzhi, Ezhil Lang etc as source code in github. Few sites work and many are not. If we have a common portal with samples installed, will be easy to explore. Need a portal like http://www.opensourcecms.com/. Need a section to list all fonts with sample images.

5. Cant use with major software
   Few Tools work as standalone. But the real need would to be used as plugin to any other major system. Example: Spellchecker by Thamizha. It works with Mozilla browser. But the needs is to work with MS Office/LibreOffice/PageMaker/InDesign. There may be tech issues like proprietary software may not allow plugins to be developed by others.

6. Lack of Graphical User Interfaces (GUI)

   UI should be nice and easy to easy adoption. Most developers are not good UI designers, and the applications are not looking good. Example - Tamil tools by Velmurugan Subramanian. This is good in linguistic features, but not good at UI.

7. Lack of user/developer documentation

   With no or less documentation, users find it is difficult on using a software further. When a developer is left with no developer documentation, he/she looks for another project with nice developer doc. Example : Velmurugan Subramanian's Tamil application is good. But its documentation is missing with the source code. The site for doc is not working now.

8. Lack of marketing done by developers

   The applications are developed and source is pushed to github. There no announcements in public mailing lists like FreeTamilComputing or similar. There may be few Facebook posts or tweets. But to get more people's attention, the project releases should be announced to wider audience.

9. Lack of Offline events

   The offline events like intro talks on Linux users groups, hackathons will inspire local people to contribute. These kind of events are not happening.

10. Not mobile friendly

    Most of the web apps wont work well with mobile browsers. Cant make mobile apps with webview. They should have mobile friendly CSS styles. Possible applications should be converted as mobile apps too. Like tamil games, spell checkers, TTS, OCR etc, should be available as mobile apps. Example: Thamizha's spellcheckers wont work with mobile browsers.

11. No lobbying

    Cant convince big giants to install the TOSS as inbuilt. Example: e-kalappai can be a part of windows by default. Indic-keyboard can be a part of android. Still these are not happened. We don't know how to do this and whom to contact.

12. No funding.
    There are no private/govt agencies for funding. No events like GSOC. No events/prizes/goodies for developers.

13. Transliteration is cheap alternative
    Institutional users of Tamil – mainly Tamil cinema industry – does not strongly patronize Tamil software in script writing, re-recording, dubbing, music production/playback singing etc. and chooses to use transliterated Tamil. However there are problems in Tamil prosody, diction which is appalling in Tamil cinema industry today, but more importantly they fail to be a market for Tamil software.

14. Lack of reusing
    Most of the features are buried with the applications itself. No libraries/API are provided.

15. Other basic issues for end-user:

    a) He/She don't know tamil typing.
    b) Tamil keyboards (physical) are not available; most of them are onscreen keyboards. keyboard covers with tamil letters embossed are available
    c) Too many keyboard layouts like Tamil Typewriter, Tamil99 cause confusion to user
    d) Too many fonts/engines. Still unicode is not a standard. Introduction of TACE/16 makes it complex especially without open-source tools for TACE encoding.
    e) Fear of typing wrongly as no default spellcheckers are available across the OS – Tamil writing is more self-critical and acts negatively to suppress Tamil usage online.

We think addressing one or more of these issues highlighted above will create a positive feedback growth-cycle for the Tamil Open-Source landscape.

# Recommendations

We feel the following recommendations are necessary to groom young talents into Tamil software generally and open-source Tamil computing in particular:

1. A central FAQ about accessing Tamil script/text functions in various programming languages to address developer training
2. There is a need for institutional effort to have a Tamil coding school
3. There is a need for Tamil marketplace for software

# Conclusions

We report in this paper, Tamil open-source software community is a vibrant place with software developers, font designers, translators, voice-over artists, and general user testers, who come together for love of their language, and promotion of critical thinking, and modern language usage in Tamil. We identify a need for institutional support at various stages from grooming software developers in Tamil, to marketing platform for Tamil software. There is bright future for tamil software if we will meet challenges it brings with it.